\documentclass[%
 aip,
 amsmath,amssymb,
 reprint,%
]{revtex4-1}

\usepackage{graphicx}% Include figure files
\usepackage{dcolumn}% Align table columns on decimal point
\usepackage{bm}% bold math
\usepackage[mathlines]{lineno}% Enable numbering of text and display math
\usepackage{braket}                        

\usepackage[utf8]{inputenc}
\usepackage[T1]{fontenc}
\usepackage{mathptmx}
\usepackage{xspace}                                            
\usepackage{color,soul}

\newcommand*{\Eh}{$E_\mathrm{h}$\xspace}

\newcommand{\cre}[1]{\hat{a}_{#1}^{\dagger}}
\newcommand{\ann}[1]{\hat{a}_{#1}}
\begin{document}

\newcommand{\boldmat}[1]{ \boldsymbol{\mathsf{#1}}}

\title{Time Dependent Adaptive Configuration Interaction Applied to Attosecond Charge Migration}

\author{Jeffrey B. Schriber}
\affiliation{Center for Computational Molecular Science and Technology, School of Chemistry and Biochemistry, Georgia Institute of Technology, Atlanta, Georgia, 30318, USA}
\email{jschriber7@gatech.edu}

\author{Francesco A. Evangelista}
\affiliation{Department of Chemistry and Cherry L. Emerson Center for Scientific Computation, Emory University, Atlanta, Georgia, 30322, USA}
\email{francesco.evangelista@emory.edu}

\date{\today}% It is always \today, today,

\begin{abstract}
In this work, we present a time-dependent (TD) selected configuration interaction method based on our recently-introduced adaptive configuration interaction (ACI).
We show that ACI, in either its ground or excited state formalisms, is capable of building a compact basis for use in real-time propagation of wave functions for computing electron dynamics.
TD-ACI uses an iteratively selected basis of determinants in real-time propagation capable of capturing strong correlation effects in both ground and excited states, all with
an accuracy---and associated cost---tunable by the user.
We apply TD-ACI to study attosecond-scale migration of charge following ionization in small molecules.
We first compute attosecond charge dynamics in a benzene model to benchmark and understand the utility of TD-ACI with respect to an exact solution.
Finally, we use TD-ACI to reproduce experimentally determined ultrafast charge migration dynamics in iodoacetylene. 
TD-ACI is shown to be a valuable benchmark theory for electron dynamics, and it represents an important step towards accurate and affordable time-dependent multireference methods.
\end{abstract}

\maketitle 

Recent developments of high-intensity attosecond laser pulses\cite{Krausz:2009hz} can provide valuable insight to the phenomena of charge migration, 
defined as the electronic motion, e.g. following ionization, that occurs before a nuclear response.\cite{Nisoli:2017hr,Leone:2014uu,Kraus:2018eo,Despre:2018et}
Charge migration has been used to explain site-selective reactivity in electronically excited peptides\cite{Weinkauf:1996ky,Weinkauf:1997jx}
and can potentially be used to direct chemical reactions into normally inaccessible pathways.
While initial work on experimental measurement and control of charge migration is promising,\cite{Kraus:2015hl} theoretical techniques are 
required to understand specific electron dynamics pathways and to begin answering questions on the greater feasibility of charge migration controlled chemistry.

Cederbaum and Zoberly first introduced the idea that electron correlation in populated excited cationic states drives attosecond charge migration following ionization.\cite{Cederbaum:1999ir,Hennig:2005dl,Breidbach:2005fd,Kuleff:2014gw,Kuleff:2016cx}
Following this work, numerous studies using the non-Dyson intermediate-state representation of the third or fourth order algebraic diagrammatic construction [ADC(3), ADC(4)],\cite{Wormit:2014kn,Dreuw:2014hd, Kuleff:2007gs,Lunnemann:2008hw,Lunnemann:2008ff,Kuleff:2009bf,Lunnemann:2009ev,Kuleff:2010km,Despre:2015jf,Golubev:2017ft,Despre:2018et}
time-dependent density functional theory (TD-DFT),\cite{Lopata:2011ce,Lopata:2012ck,Bruner:2017jd} and time-dependent density matrix renormalization group (TD-DMRG)\cite{Frahm:2019bs,Baiardi:2019fb} 
have been employed to further understand pure electron dynamics following ionization.
The major theoretical challenge of describing coherent electron dynamics following ionization is in accurately characterizing the populated cationic states.
ADC(3), ADC(4), and TD-DFT are affordable options to study charge migration, but they inherently depend on a single-reference description of correlation effects. 
Failure to characterize strong correlation in excited states can lead to a qualitatively incorrect description of the dynamics and is likely to occur when
numerous near-degenerate cationic states with many coupled excitations become populated.
TD-DMRG can treat an electronic state beyond the traditional excitation level hierarchy and is thus well-positioned for computing complicated dynamics, 
despite challenges associated with excited-state DMRG algorithms and propagating a matrix product state wave function.\cite{chan2002highly,Ren:2018kg,Frahm:2019bs,Baiardi:2019fb} 

In this work, we generalize the adaptive configuration interaction\cite{schriber2016communication,Schriber:2017jd,Schriber:2018hw} method (ACI) to a time-dependent version (TD-ACI) for simulating electron dynamics.
ACI is based on the very old idea of selected CI\cite{Gershgorn:1968go,Pipano:1968gi,Bender:1969he,Whitten:1969hg,Huron:1973cb,Buenker:eo,Buenker:1975hr,evangelisti1983convergence,Cimiraglia:1985eq,harrison1991approximating,Hanrath:1997tm,Stampfuss:2005bk,bytautas2009priori,Greer:1995fm,Greer:1998ew}
and is one of numerous new manifestations of these techniques.\cite{coe2012development,coe2013state,knowles2015compressive,McClean:2015tz,bunge2006selected,evangelista2014adaptive,tubman2016deterministic,Ohtsuka:2017cw,Giner:2013cg, giner2016quantum,holmes2016heat,Sharma:2017iu,Holmes:2017gp,Li:2018gj}
TD-ACI can be viewed as an approximation to time-dependent complete active space (CAS) techniques, which have been successfully applied to photoionization processes,\cite{Klinkusch:2009iw,Hochstuhl:2012hu,Bauch:2014iv,Goings:2017bp,Peng:2018fn,Liu:2019fn} 
where determinant screening in ACI enables the use of much larger CAS spaces.
Due to its systematic improvability and demonstrated ability to treat many strongly correlated electrons in ground and excited states, 
TD-ACI will be essential for benchmarking time dependent methods 
and for providing insight to the role of electron correlation in attosecond electron dynamics.

\begin{figure*}[ht!]
\centering
\includegraphics[width=\textwidth]{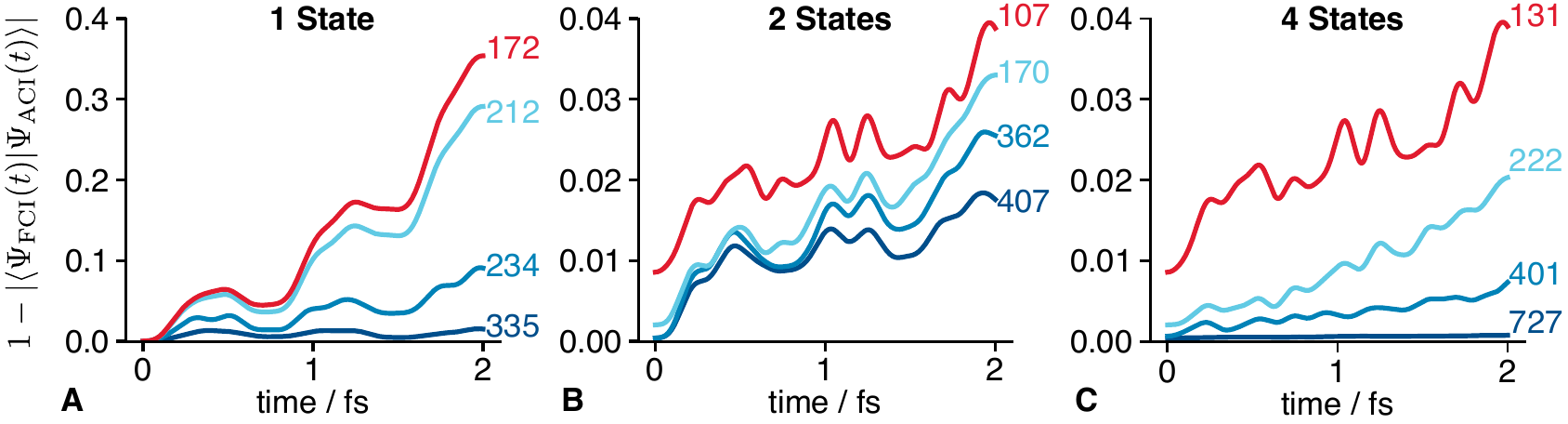}
\caption{Errors in the overlap between TD-ACI wave functions and the exact wave function, for which $|M^{N-1}|=3920$ determinants.
TD-ACI wave functions are computed with various $\sigma$ values using one ($\textbf{A}$) , two ($\textbf{B}$) , or four ($\textbf{C}$)  reference states in computing the cationic basis.  
Each error curve is labeled with the number of determinants in the cationic space. The FCI wave functions are propagated exactly, 
and TD-ACI wave functions are propagated using RK4 with a timestep of 0.05 as.}
\label{fig:benzene_ovp}
\end{figure*}

Studying ultrafast dynamics of electrons requires the solution to the time-dependent Schr\"odinger equation (TDSE), in atomic units,
$i\frac{\partial}{\partial t}\ket{\Psi(t)} = \hat{H}\ket{\Psi(t)}$,
which dictates how a wave function, $\Psi(t)$, evolves in time according to the Hamiltonian ($\hat{H}$).
For a time-independent $\hat{H}$, the evolution of a wave function can be written exactly,
\begin{equation}
\ket{\Psi(t + \Delta t)} = e^{-i\hat{H}\Delta t}\ket{\Psi(t)},\label{eq:eprop}
\end{equation}
where $\Delta t$ is the timestep over which the wave function evolves.
To avoid the combinatorial complexity of the exact full configuration interaction (FCI) solution to the TDSE,
many approximate time-dependent methods in electronic structure theory have been proposed to study pure electron dynamics,
including real-time versions of Hartree--Fock (HF) theory,\cite{Kulander:1987ky,Li:2005fw} 
configuration interaction,\cite{Rohringer:2006kp,Schlegel:2007ke,Krause:2007dm,Krause:2014io,Sato:2015jh,Peng:2018fn,Ulusoy:2018hh} 
multiconfigurational self consistent field wave functions,\cite{Meyer:1990if,Nest:2005fu,Miranda:2011id} 
density functional theory,\cite{Theilhaber:1992gr,Yabana:1996fb,Baer:2001dg,Lopata:2011ce}
and coupled-cluster theories.\cite{Dalgaard:1983ky,Huber:2011iv,Kvaal:2012ht,Nascimento:2016ca,Nascimento:2017fx}
To study charge migration following ionization, equation \ref{eq:eprop} needs to be accurately approximated by ($i$) generating an accurate initial ionized state, 
($ii$) building a tractably-sized cationic Hamiltonian that well-describes all populated cationic states throughout propagation, and ($iii$) avoiding evaluation
of the exponential using conventional numerical integrators.

We use ACI to build a compact determinantal representation of $\hat{H}$ and a well-defined initial ionized state.
ACI uses an iterative screening algorithm to build a model space, $M^{N}$, composed of $N$-electron Slater determinants ($\Phi^{N}$), such that the resultant
wave function, $\ket{\Psi^{N}} = \sum_{\Phi^{N}_{\mu}\in M^{N}}\ket{\Phi_{\mu}^{N}}c_{\mu}$, produces an energy error approximately equal to a user-defined parameter, $\sigma$ ($|E_{\rm{ACI}} - E_{\rm{FCI}}| \approx \sigma$).\cite{schriber2016communication}
We have reported several algorithms to compute excited states with ACI, 
including a state-averaged ACI (SA-ACI) which optimizes a single model space by defining determinant importance as the contribution to the correlation energy averaged
over several roots.\cite{Schriber:2017jd} 
This algorithm is very useful in formulating a time-dependent theory as it produces a compact Hamiltonian with a controllable average energy error over any number of roots. 

To study charge migration following ionization, we first perform a SA-ACI computation on the lowest one or few states of the neutral molecule of interest
to build the ground state wave function, $\ket{\Psi^{N}_{0}}$.
We invoke the sudden ionization approximation,\cite{Cederbaum:1999ir,Despre:2018et,Frahm:2019bs}
where the initial wave function for the simulation [$\Psi^{N-1}(t=0)$] is defined immediately 
after the ionization process by annihilating an electron in spin orbital $\phi_{i}$ and normalized appropriately as
\begin{equation}
\label{eq:initial}
\ket{\Psi^{N-1}(t=0)} = \frac{\hat{a}_{i}\ket{\Psi^{N}_{0}}}{\bra{\Psi_{0}^{N}}\hat{a}^{\dagger}_{i}\hat{a}_{i}\ket{\Psi_{0}^{N}}}.
\end{equation}
The sudden ionization approximation provides a well-defined initial state suitable for benchmarking the propagation in TD-ACI field-free with a time-independent Hamiltonian.
In practice, TD-ACI simulations can include the ground electronic state interacting with an ionizing pulse, but invoking the sudden ionization approximation allows us to effectively decouple errors
associated with TD-ACI and those connected to our choice of initial conditions.
Since ACI uses a linear expansion of Slater determinants, the initial state can be defined by either a localized hole or a superposition of holes without complications possible in other approaches.\cite{Eshuis:2009ep,Maitra:2001hz,Bruner:2017jd}

We then define the basis for the cationic Hamiltonian, $M^{N-1}$, as the set of cationic determinants, $\{\Phi^{N-1}\}$, generated by applying a single annihilation over all spin orbitals to 
all determinants in the original basis $M^{N}$, $M^{N-1} = \{\Phi^{N-1}_{I}: \Phi^{N-1}_{I} = \hat{a}_{i}\ket{\Phi^{N}_{\mu}},~  \forall i \in \mathbf{A} ~\mathrm{and}~  \forall \Phi^{N}_{\mu} \in M^{N}\}$,
where $\mathbf{A}$ denotes the set of all active orbitals. 
Lastly, we define the TD-ACI wave function with the time-dependence entirely encoded in the expansion coefficients [$c_{I}(t)$] as
$\ket{\Psi^{N-1}(t)} = \sum_{\Phi^{N-1}_{I}\in M^{N-1}}c_{I}(t)\ket{\Phi^{N-1}_{I}}.$
Starting from the initial condition [equation~\eqref{eq:initial}], the wave function $\Psi^{N-1}(t)$ can be determined at any time $t$ by integration of the TDSE.
Exact propagation via equation \ref{eq:eprop} requires complete diagonalization of the cationic Hamiltonian and is unfeasible for realistic simulations.
Instead, we use the fourth-order Runge--Kutta (RK4) algorithm with fixed timesteps,\cite{Press:2007te,GomezPueyo:2018ez,Ren:2018kg,Frahm:2019bs}
which for a time-independent $\hat{H}$ is equivalent to approximating the exact exponential propagator using a Taylor series truncated to fourth order. 
We monitor the migration of the ionized hole throughout the molecule using the 
hole occupation number, $n_{i}$ for an orbital $i$, defined as the difference between the occupation of orbital $i$ in the ground state and the ionized state at time $t$,
$n_{i}(t) = \bra{\Psi^N_0}\cre{i}\ann{i}\ket{\Psi^N_0} - \bra{\Psi^{N-1}(t)}\cre{i}\ann{i}\ket{\Psi^{N-1}(t)}$.

We first study dynamics triggered by valence ionization in benzene to understand the effects of using a truncated cationic model space and an approximate time propagator.
The dynamics of the ionized state is characterized by hole migration within the $\pi/\pi^*$ manifold, with weak hole-occupation of a nearby $\sigma$-bonding orbital.
The benzene geometry was optimized using DFT with a B3LYP functional and the cc-pVDZ basis set using the \caps{Psi4} program.\cite{Parrish:2017hga} 
ACI computations use a CAS(8,8) containing the $\pi/\pi^*$ valence space and the energetically nearest $\sigma/\sigma^*$ bonding/antibonding pair, and they
employ a restricted HF reference computed with the cc-pVDZ basis and density fitted integrals using the cc-pVDZ-JKFIT basis.\cite{Weigend:2002ga}  
The computation was run using $C_1$ symmetry, but we refer to using corresponding $D_{2h}$ labels for clarity, all plotted in figure S1.
In the TD-ACI simulation, we use a time step of 0.05 as for a total time of 2 fs, and the initial state is prepared by annihilating the alpha $1b_{1u}$ spin orbital. 

Upon ionization of the $1b_{1u}$ orbital, the hole migrates to a superposition of the degenerate $1b_{2g}$ and $1b_{3g}$ orbitals and back smoothly with a frequency of roughly 750 as.
The oscillation is faster by about 200 as compared to previously reported ADC(3) results because of the minimal CAS(8,8) employed in this work.\cite{Despre:2015jf}
This reduced model is nonetheless an effective test of our theory because the oscillating hole occupations require determinants 
different in character from those that define the ground cationic state. 
We test three different schemes to build $M^{N-1}$ from a SA-ACI computation of $M^{N}$ optimized with respect to one, two, or four roots 
of the neutral species. These time-independent computations are run with values of $\sigma$ chosen to produce similar dimensions of $M^{N-1}$ to facilitate comparisons.

With the various $M^{N-1}$ bases, we propagate the initial wave functions using the RK4 algorithm with a 0.05 as timestep, 
and we plot the error in their overlaps with respect to the FCI result in Figure \ref{fig:benzene_ovp}.
Due to the short timestep used, the observed errors are largely resultant from truncation alone, and we see it is not necessarily correlated with the number of determinants in each computation. 
For example, the propagations with $M^{N-1}$ generated from a single state (\ref{fig:benzene_ovp}.\textbf{A}) show errors in the overlap up to 0.3 at 2 fs when the cationic space contains about 200 determinants. 
With a similar dimension, the two-state variant already shows an approximate 10-fold increase in accuracy. 
This increase in accuracy is due to the influence of two-hole/one-particle states and simple hole-excited states in the hole dynamics.\cite{Despre:2015jf} 
To describe all of these states, $M^{N-1}$ needs determinants with single, double, or higher excitation character with respect to ionized ground and excited states. 
Our results indicate that such a determinantal makeup is most effectively achieved by ionizing determinants from an SA-ACI computation done with respect to several excited states.

\begin{figure}[t!]
\includegraphics[width=3.275in]{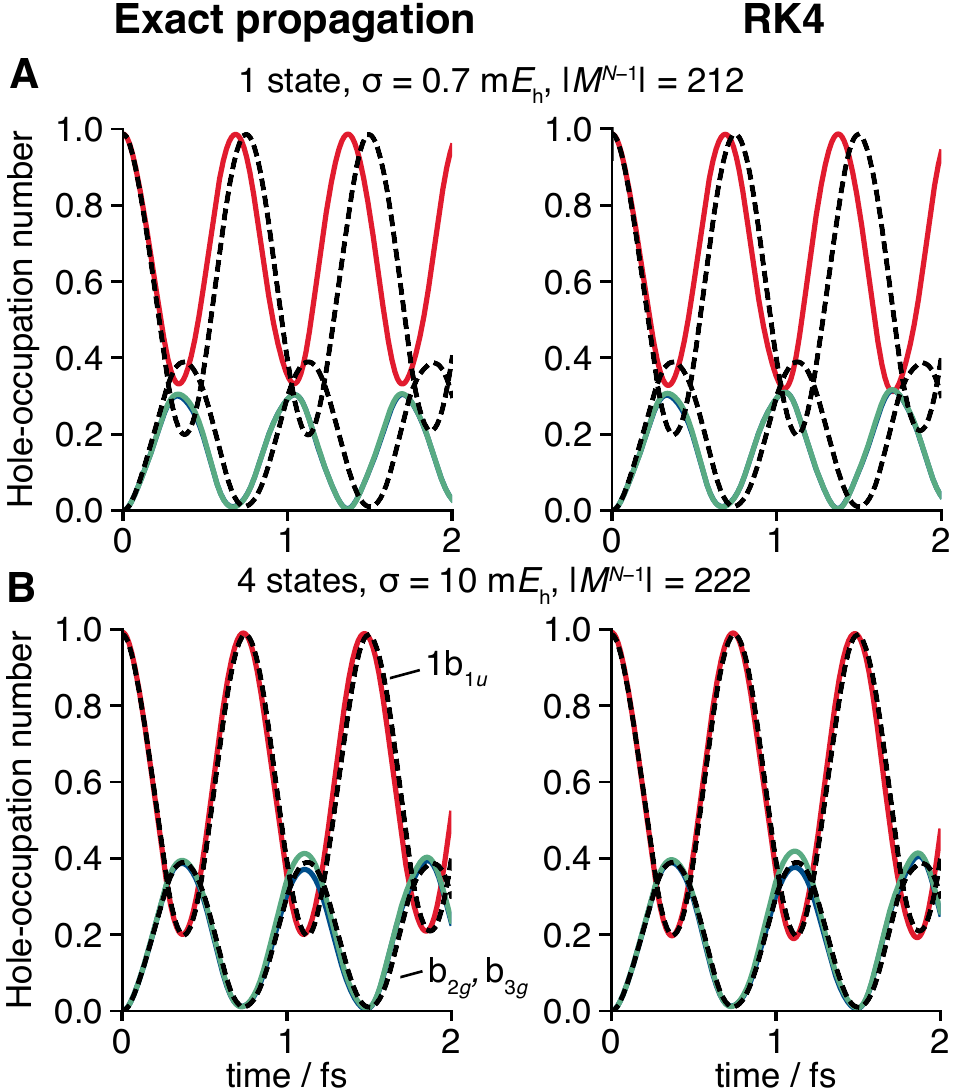}
\caption{Hole occupations of $1b_{1u}$, $1b_{2g}$, and $1b_{3g}$ orbitals computed using TD-ACI with $M^{N-1}$ optimized for 1 ($\textbf{A}$) and 4 ($\textbf{B}$) states using
exact propagation (left) and RK4 propagation (right) with a 0.05 as time step.
TD-FCI occupations are shown withe a dashed grey line.}
\label{fig:benzene_occ}
\end{figure}

The hole migration from the $1b_{1u}$ orbital to a superposition of $1b_{2g}$ and $1b_{3g}$ orbitals is also shown in Figure \ref{fig:benzene_occ}. 
%The exact hole occupation of the $1b_{1u}$ orbital is shown in each plot starting at unity, and the  $1b_{2g}$ and $1b_{3g}$ hole occupations start from zero and are always identical. 
In the left plots, we show hole occupations computed with TD-ACI using the exact propagator with either one (\ref{fig:benzene_occ}.\textbf{A}) or four (\ref{fig:benzene_occ}.\textbf{B}) optimized roots in defining $M^{N-1}$. 
The plots on the right show the same data, but instead using the RK4 propagator.
Our first observation is that the TD-ACI dynamics are indistinguishable when using either exact or RK4 propagatiors, showing that
our expected propagation errors are negligible even with determinant selection so long as an appropriate time step is chosen.
When the $M^{N-1}$ is optimized with respect to one root, the hole occupation oscillates too quickly and does not transfer from the $1b_{1u}$ orbital (red) to the $1b_{2g}$ and $1b_{3g}$ orbitals
(blue and green) with enough magnitude. Interestingly, the hole occupations of the $1b_{2g}$ and $1b_{3g}$ orbitals are correctly always identical.
The four-state procedure to build a similarly-sized $M^{N-1}$ gives occupations nearly indistinguishable from the exact result. 
These results indicate that TD-ACI is a viable technique in studying charge migration resultant from ionization, particularly if the basis for the cationic Hamiltonian is truncated 
using importance criteria that can consider multiple roots. This use of an SA-ACI computation to build the basis enables a faster convergence to the $\sigma=0$ limit,
and does not bias the determinantal makeup of cationic Hamiltonian towards the ground cationic state. 

\begin{figure*}[t!]
\centering
\includegraphics[width=\textwidth]{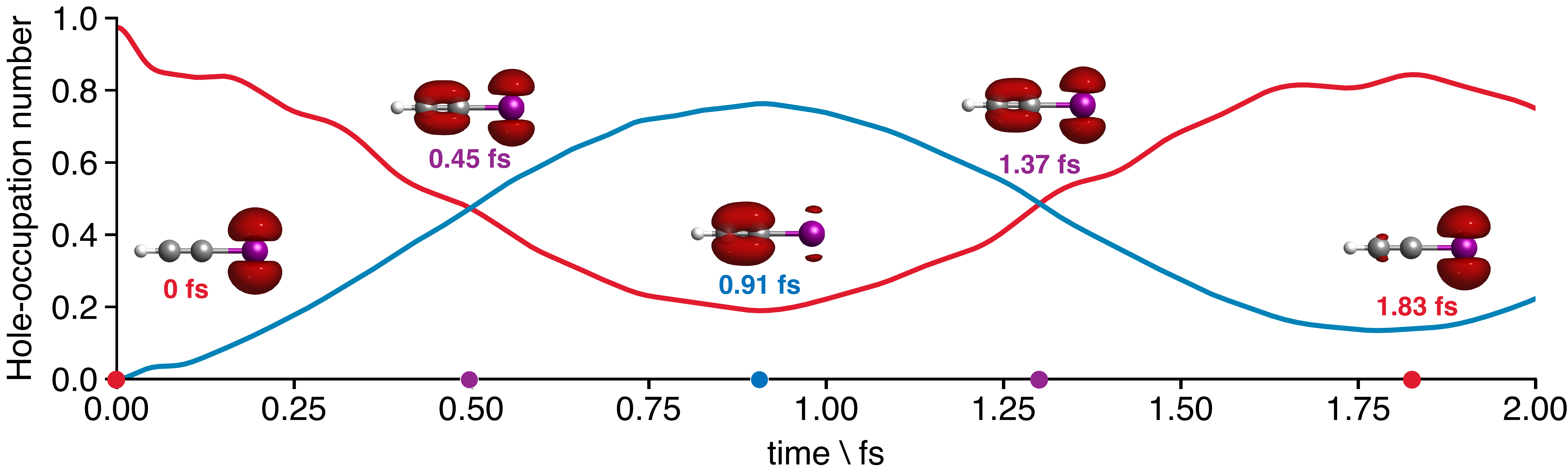}
\caption{Hole occupations of the $5p$-like orbital on iodine (red) and the $\pi$ orbital on the acetylene (blue) in the same plane. 
Evolution of the hole density ($\rho_\mathrm{h}$), as defined in the text, is shown at maxima, minima and intersections of the hole occupation curves.}
\label{fig:hcci_8}
\end{figure*}
Iodoacetylene is a valuable model of charge migration for theorists and experimentalists alike. Recently, Kraus \textit{et. al} have used high-harmonic	
generation to achieve 100 as resolution and control over charge migration following ionization.\cite{Kraus:2015hl,Kraus:2018eo}
The dynamics following ionization was also investigated theoretically in a TD-DMRG study using a CAS(16,36) active space.\cite{Frahm:2019bs}
In iodoacetylene, two cationic states drive the dynamics, one characterized by a hole in the $5p$-like orbitals of iodine perpendicular to the molecular axis, and the
other having a hole in the two $\pi$-bonding orbitals in the acetylene triple bond. 
These states are near degenerate and a multiconfigurational time-dependent approach may be required.
The dynamics of ionized iodoacetylene involve the hole migrating between the iodine
and acetylene groups with an experimentally measured frequency of about 0.93 fs.\cite{Kraus:2015hl,Kraus:2018eo} 
Our final test of TD-ACI is to reproduce this migration frequency using a truncated basis of cationic determinants. 

We optimize the geometry with DFT using a PBE0\cite{Perdew:1996iq} functional and def2-SVP basis set.\cite{Weigend:2005vs} 
This basis was used in all iodoacetylene computations in addition to an effective core potential used to remove 28 core electrons.
 Within the remaining 59 orbitals, we use a CAS(16,22) for the neutral species, which includes 
all molecular orbitals generated from the $2p$-like orbitals in the acetylene group, and $4d$ and $5p$-like molecular orbitals from the iodine.
In the dynamics simulations, we use a timestep of 0.02 as
for a total simulation time of 2 fs. The initial state is prepared by annihilating a valence $5p$-like orbital on the iodine atom, and all computations are run in $C_1$ symmetry
with split-localized orbitals in the active space.

To analyze the charge migration dynamics, we compute a representation of the density of the ionized hole ($\rho_\mathrm{h}$) by scaling the HF orbitals
by the corresponding hole occupation,\cite{Cederbaum:1999ir,Frahm:2019bs}
$\rho_\mathrm{h} = \sum_{i}|\phi_i|^2n_{i}$.
Our first simulation uses a two-state SA-ACI computation with $\sigma = 10 ~m$\Eh to generate a cationic basis containing 244,361 determinants,
where the cationic Hilbert space contains roughly $5\times10^{10}$ determinants. 
We plot the hole occupation numbers for the ionized $5p$-like orbital on the iodine and the in-plane $\pi$-bonding orbital of the acetylene in Figure \ref{fig:hcci_8} for the 2 fs simulation. 
The hole migrates from the iodine to the acetylene and back in about 0.91 fs, agreeing closely with the experimental frequency of 0.93 fs. 
The migration still leaves some degree of hole occupation on each moiety, which we show Figure \ref{fig:hcci_8}. 
The relatively good agreement between our computed frequency of hole migration and the experimental value suggests
that the ACI procedure is building an adequate space of determinants to describe the relevant cationic states.

To test the sensitivity of the dynamics with respect to $M^{N-1}$, 
we show hole migration times with TD-ACI using cationic spaces built from one, two and three root SA-ACI computations
in Table \ref{tab:st_comp}.
The acetylene migration time is defined as the time required for the ionized hole, initially on the iodine, to maximally populate on the acetylene group.
The iodine migration time is the time elapsed from initial ionization to repopulation of the hole density on iodine after it has migrated.
For a nearly constant size of $M^{N-1}$, we see that increasing the number of states used to 
generate the cationic basis has a negligible effect on the hole migration times. Only when the total number of cationic determinants is increased, even when optimized for the neutral
ground state, do we see the migration times approach the experimental and TD-DMRG values. This result depends solely on the determinantal makeup of the cationic states most
important in forming the evolving wave packet. For iodoacetylene, these states are simple 1-hole cationic states with the hole located on the iodine or acetylene moieties, and no coupled
electronic excitations are as significant as they were for our previous study on benzene. As a result, the dominant contribution for both hole states comes from different annihilations 
of the ground state and not from annihilations of excited states. 

Our correlation treatment provides accurate dynamics despite the neglect of dynamical correlation.
While fortuitous cancellation of error is possibly present and some dynamical correlation may be recovered within our active spaces, 
the success of the active-space treatment suggests that dynamical correlation is relatively unimportant in accurately defining the relevant cationic states for propagation.
While dynamical correlation is likely necessary for accurately computing other time-dependent properties, TD-ACI shows great promise in computing time-dependent reference wave functions.
A very appealing property of our TD-ACI approach is that we can effectively optimize $M^{N-1}$ regardless of the character of the relevant cationic states, whether simple or complex.

\begin{table}[t!]      
    \centering          
    \footnotesize    \caption{Iodoacetylene hole migration times computed with TD-ACI using various $M^{N-1}$, and from experiment (Exp.) from Ref.~\citenum{Kraus:2015hl}.}
    \begin{tabular*}{3.275in}{@{\extracolsep{\fill}} crcc}   
   
    \hline  
                     
    \hline                  
     & & \multicolumn{2}{c}{Migration Time (as)} \\ \cline{3-4}
    Number of States & $|M^{N-1}|$ & Acetylene & Iodine  \\   \hline                            
     1 & 10127 & 862 & 1684 \\
     2 & 10827 & 864 & 1660 \\
     3 & 10759 & 868 & 1610 \\
     1 & 93554 & 886 & 1780 \\    
     2 & 244361 & 907 & 1828 \\
    Exp. & & 930 & 1850 \\                                                  
                                                   
   \hline 
                                                    
   \hline
   \end{tabular*} 
   \label{tab:st_comp}                                                                                                                               
\end{table} 

In this work, we have extended the ACI approach to simulate ultrafast electron dynamics. TD-ACI is able to effectively model the real-time propagation
of the exact wave function using a truncated space of determinants selected by the time-independent ACI algorithm. We apply this methodology to charge migration dynamics
that follow ultrafast ionization. We find that ACI is well-suited to find determinants relevant to the initial state, and determinants important at later points in time can be initially identified 
in excited state computations to reduce systematic increases in error as the wave function evolves. We also find that the single-state ACI computation 
can recover these determinants if a sufficiently large value of $\sigma$ is used. Due to the short time scale of charge migration, we did not encounter significant issues in using approximations
to the exact time propagator, though these effects could be important in longer simulations. Propagation error and truncation errors were studied using a charge migration model in benzene,
where we found that cationic spaces optimized for multiple roots are needed to efficiently capture dynamics in which two-hole one-particle states are relevant. 
This type of state is not always relevant to the dynamics in propagating our initial state, as we saw in our application to hole migration in iodoacetylene. Using various schemes to build our 
cationic basis, we were able to compute hole migration times between acetylene and iodine groups that match experimental results. 

We anticipate numerous future directions of study. While the SA-ACI procedure was successful in building a determinantal
space for the entire dynamics, we envision an even more efficient scheme where the selection and removal of determinants can be done during the simulation itself. For example,
one can imagine an algorithm that estimates the importance of a determinant at a future timestep using low-order approximate propagation. Additionally, integration of the current TD-ACI
scheme with a time-dependent Hamiltonian would enable \textit{ab initio} studies of molecules interacting with fluctuating electric or magnetic fields. Finally, we envision including
dynamical correlation effects beyond our active space treatment by combining an effective Hamiltonian theory with our time-dependent reference wave functions.

\begin{acknowledgments}
This work was supported by the U.S. National Science Foundation under Grant No. CHE 1900532 and a Camille Dreyfus Teacher--Scholar Award (TC-18-045). J.B.S. would like to thank Dr. Tianyuan Zhang for helpful discussions on approximate propagation techniques.
\end{acknowledgments}

\end{document}